\documentclass[aps,prb,reprint, superscriptaddress]{revtex4-2}

\usepackage{hyperref}
\usepackage{graphicx}
\usepackage{amsmath}
\usepackage{amssymb}
\usepackage{amsfonts}
\usepackage{braket}
\usepackage{color}
\usepackage{bm}

\renewcommand{\Im}{\operatorname{Im}}

\newcommand{\up}{\uparrow}
\newcommand{\dn}{\downarrow}

\begin{document}
\title{Floquet prethermalization and Rabi oscillations in optically excited Hubbard clusters}

\author{Junichi Okamoto}
\affiliation{Institute of Physics, University of Freiburg, Hermann-Herder-Str.
3, 79104 Freiburg, Germany}
\affiliation{EUCOR Centre for Quantum Science and Quantum Computing, University
of Freiburg, Hermann-Herder-Str. 3, 79104 Freiburg, Germany}
\author{Francesco Peronaci}
\affiliation{Max Planck Institute for the Physics of Complex Systems, Dresden
01187, Germany}
\date{\today}

\begin{abstract}
We study the properties of Floquet prethermal states in two-dimensional Mott-insulating Hubbard clusters under continuous optical excitation. With exact-diagonalization simulations, we show that Floquet prethermal states emerge not only off resonance, but also for resonant excitation, provided a small field amplitude. In the resonant case, the long-lived quasi-stationary Floquet states are characterized by Rabi oscillations of observables such as double occupation and kinetic energy. At stronger fields, thermalization to infinite temperature is observed. We understand these results in terms of the optically driven two-site Hubbard model and by means of time-dependent perturbation theory.
\end{abstract}

\maketitle

\section{Introduction}

Coherent optical manipulation of matter is a growing field of study due to the development of intense laser sources. Photoinduced phase transitions in many-body systems are now possible with non-thermal processes without loss of quantum nature \cite{averitt2002, yonemitsu2008, giannetti2016, nicoletti2016, citro2018, ishihara2019}. For example, recent mid-infrared pump-probe experiments revealed exciting phenomena such as light-induced superconductivity \cite{fausti2011, hu2014, mitrano2016}, ultrafast structural transitions \cite{tobey2008, rini2007}, and metastable charge ordering \cite{stojchevska2014}, which are driven by short optical excitation of phonons. More direct manipulation of electronic states can be realized through the so-called Floquet engineering by continuous optical excitation \cite{bukov2015a, eckardt2017, oka2019}. Notable achievements in this direction are the prediction and realization of dynamical localization \cite{dunlap1986, aguado1997, ishikawa2014} and topological band structures \cite{oka2009, kitagawa2011, lindner2011, mciver2020}.

In the high-frequency limit, Floquet theory provides a good description of low-energy phenomena in terms of effective static Hamiltonians. In this limit, heating is a rather minor effect even in interacting systems, since the drive frequency is away from any characteristic absorption energy of the system \cite{abanin2015, mori2016, abanin2017}. Thus, long-lived quasi-stationary states, so-called Floquet prethermal states (FPSs), are realized \cite{messer2018, sandholzer2019}. On the other hand, when the drive frequency is near an energy scale of a generic interacting system, or its submultiples, heating is expected due to possible photon absorption processes. In isolated systems, this leads to thermalization to infinite temperature \cite{dalessio2014}. Thus, resonant excitation makes the analysis by the Floquet picture more complicated.

As an alternative to the high-frequency limit, Floquet prethermalization is also observed in systems close to integrability, in which quasi-integrals of motion constrain the dynamics for finite but long times \cite{oka2008,poletti2011,dalessio2013, citro2015,  bukov2015, canovi2016, herrmann2017, weidinger2017, mori2018, seetharam2018, peronaci2018, peronaci2020}. In particular, systems in the Mott-insulating phase were shown to display extremely long-lived prethermal states, even for frequencies close to resonance \cite{herrmann2017,peronaci2018}. Further investigation of the stability and controllability of these long-lived quasi-steady states is of great importance to advance the Floquet engineering protocols for generic frequencies.

In this work, we study Floquet prethermalization in two-dimensional (2D) Hubbard clusters under continuous optical excitation. The setup can be realized in solid-state systems such as quantum dot arrays under laser fields \cite{oosterkamp1998, cole2001, vion2002, creffield2002} or in ultracold atoms in shaking optical lattices~\cite{desbuquois2017, messer2018, sandholzer2019}. Starting from the Mott-insulating state, we calculate the driven time evolution by exact diagonalization. We find that Floquet prethermal states emerge even at frequencies resonant with absorption-peak energies, provided a small field amplitude. Remarkably, these long-lived quasi-steady states show Rabi oscillations of observables such as double occupation and kinetic energy. The spectral density shows that the system oscillates between the ground state and the one-photon excited state. For stronger excitation, the system goes to the infinite-temperature state, in general, and the spectral density is spread over many excited states. We elucidate the origin of the Rabi oscillations of the FPSs with the aid of time-dependent perturbation theory, and by considering the two-site driven Hubbard model.

The remainder of the paper is organized as follows. Section \ref{sec:formalism} introduces the model that we consider and summarizes the method that we employ. In Sec.~\ref{sec:two} we illustrate the  physics of Rabi oscillations in a driven two-site Hubbard model. Several observables to characterize the time-evolved states are introduced. Section \ref{sec:2d} shows the results on the two-dimensional Hubbard cluster, together with the time-dependent perturbation analysis. Section \ref{sec:conclusions} presents our conclusions. Detailed results on the drive amplitude dependence are given in Appendix~\ref{sec:appA}.

\section{Formalism}\label{sec:formalism}

\begin{figure}
\includegraphics[width=\columnwidth]{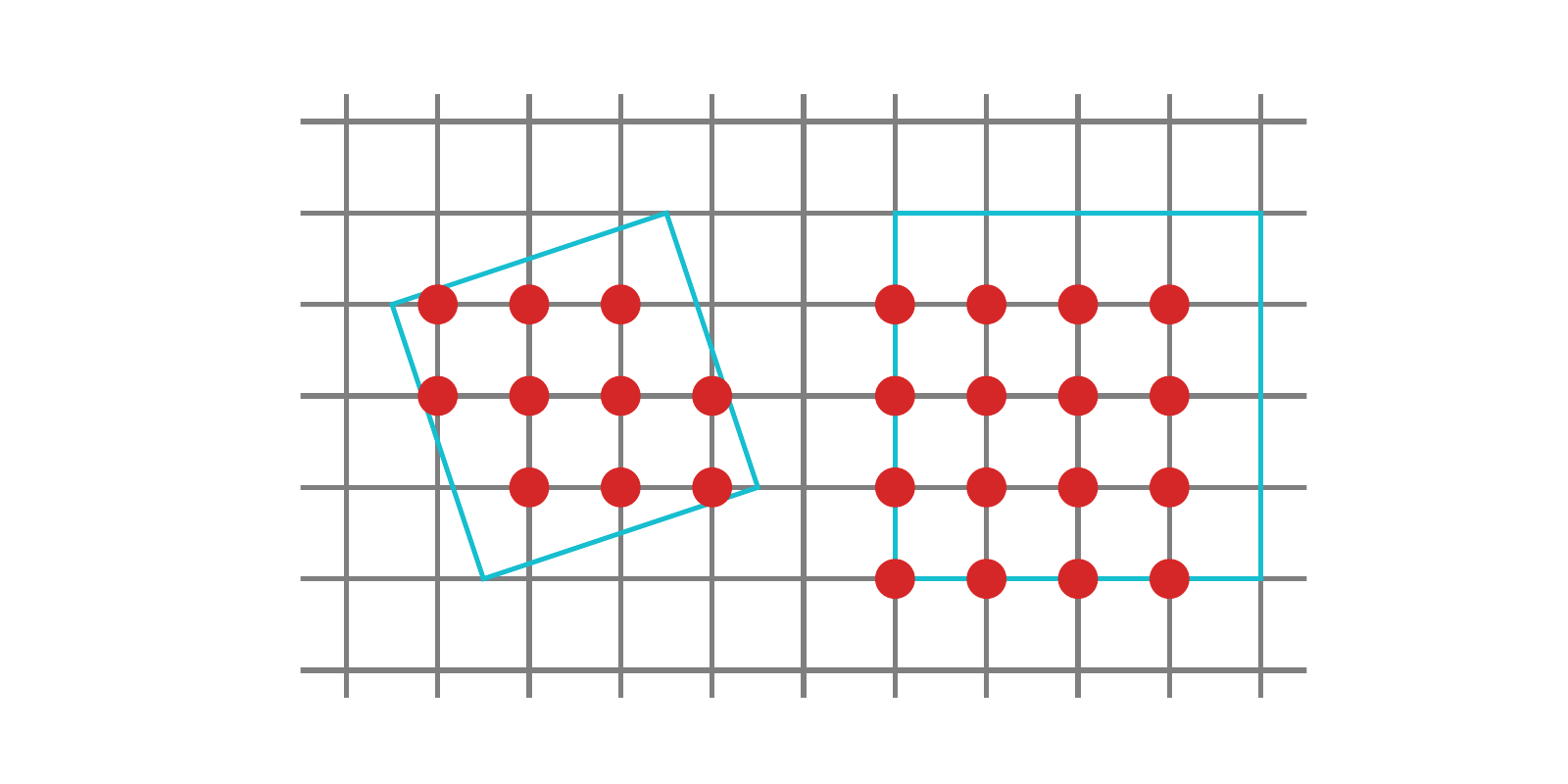}
\caption{\label{fig:lattice} Two-dimensional cluster geometries. In this work, we consider mostly the cluster with $L=10$ sites (left), while we use the cluster with $L=16$ sites (right) to check the results.}
\end{figure}

We study a half-filled Hubbard model in two dimensions under continuous optical excitation, 
\begin{equation}
\begin{split}
H(t) &= H_J (t) + U \sum_i n_{i\up} n_{i\dn} ,\\
H_J(t) &= - \sum_{\braket{i,j}, \sigma}J_0 e^{-i\frac{e}{\hbar}\int_{\bm{R}_{j}}^{\bm{R}_{i}} d\bm{r} \cdot \bm{A}(t)} c^{\dagger}_{i \sigma}c_{j\sigma} + \text{H.c.}\\
&= - \sum_{\braket{i,j}, \sigma}J_0 e^{-i \frac{e}{\hbar}  \bm{l}_{ij} \cdot \bm{A}(t)} c^{\dagger}_{i\sigma}c_{j\sigma} + \text{H.c.},
\end{split}
\label{eq:H_total}
\end{equation}
where $c_{i \sigma}$ and $c^{\dagger}_{i \sigma}$ are annihilation and creation operators at site $i$ with spin $\sigma$, and $\bm{l}_{ij} = \bm{R}_{i} - \bm{R}_{j}$ the bond vector connecting sites $i$ and $j$ at positions $\bm{R}_i$ and $\bm{R}_j$. Opposite spin densities $n_{i \sigma} = c^{\dagger}_{i\sigma} c_{i \sigma}$ are subject to a local Hubbard repulsion $U$. We take $\hbar = c = e = 1$, and use $J_0 = 1$ and $U = 6$, which ensures that the initial ground state is a Mott insulating state. The lattice geometries that we consider are given in Fig.~\ref{fig:lattice} \cite{prelovsek2013}, and periodic boundary conditions are used. Optical excitation is induced by electric fields $\bm{E}(t) = - \partial
\bm{A}(t)/\partial t$ along the $x$-axis, where the vector potential $\bm{A}(t)$ is switched on as
\begin{equation}
\begin{split}
\bm{A}(t) &= A(t) \hat{x} = A_0 \cos(\omega_d t) f(t)  \hat{x}, \\
f(t) &=
\begin{cases}
\exp \left[ - 4\log(2) \left( \frac{t-\tau_0}{\tau} \right)^2\right] & t < \tau_0 \\
1 & t \geq \tau_0
\end{cases},
\end{split}
\end{equation}
We have checked that fields along the diagonal direction $(\hat{x}+\hat{y})$ bring similar results.

The initial ground state is calculated by the conventional Lanczos method \cite{lanczos1950, Dagotto1994, weisse2008, prelovsek2013}, and the subsequent time-evolution is implemented by the Krylov-space method with time step $dt= 0.02$ and Krylov dimension $M = 20$ \cite{park1986, manmana2005, balzer2012, innerberger2020}. In the following, we focus on the cluster with $L=10$ sites (see Fig.~\ref{fig:lattice}). We have checked that results hold qualitatively unchanged with $L=16$ sites or with a one-dimensional lattice.

To characterize the time-evolved states, we calculate double occupation $d(t) = \sum_i \braket{n_{i\up}n_{i\dn}}/L$ and kinetic energy $E_\text{kin}(t) = \braket{H_J(t)}/L$. Since time evolution is unitary, isolated systems do not thermalize as a whole. For a large enough system size, however, local observables do thermalize, as described by the eigenstate-thermalization hypothesis \cite{dalessio2014}. In driven isolated systems, this leads towards the infinite-temperature state, which is characterized by $d = 0.25$ and $E_{\text{kin}}=0$. We further introduce two quantities to obtain a detailed picture of the time evolution. One is the spectral density of the wavefunction $\ket{\psi(t)}$,
\begin{equation}
S(t,\omega) = \sum_{n} \lvert\braket{n|\psi(t)}\rvert^2 \delta(\omega -\epsilon_n),
\label{eq:spectral_func}
\end{equation}
where $\ket{n}$ and $\epsilon_n$ are the eigenstates and eigenenergies (with respect to the ground state) of the zero-field Hamiltonian. Such a spectral decomposition can be approximately calculated by the Lanczos method.  The other quantity is effective dimension \cite{mori2018} or inverse participation ratio \cite{plodzien2019},
\begin{equation}
\kappa(t) = \frac{1}{\sum_n \lvert\braket{n | \psi(t)}\rvert^4},
\label{eq:kappa}
\end{equation}
which quantifies how many of the energy eigenstates are occupied. For example, in the ground state, the spectral density is peaked at zero energy and $\kappa=1$, whereas if the system oscillates between two levels we have $\kappa \approx 2$ and two peaks in the spectral density. The infinite-temperature limit may be defined as the spreading of the wavefunction over all the states of the Hilbert space $\mathbb{H}$ and thus $\kappa=\dim \mathbb{H}$. We note that Loschmidt amplitudes can be used to calculate the spectral density as well \cite{Kennes2020, maislinger2020}.

\section{Two-site model}\label{sec:two}
\begin{figure}
\includegraphics[width=\columnwidth]{{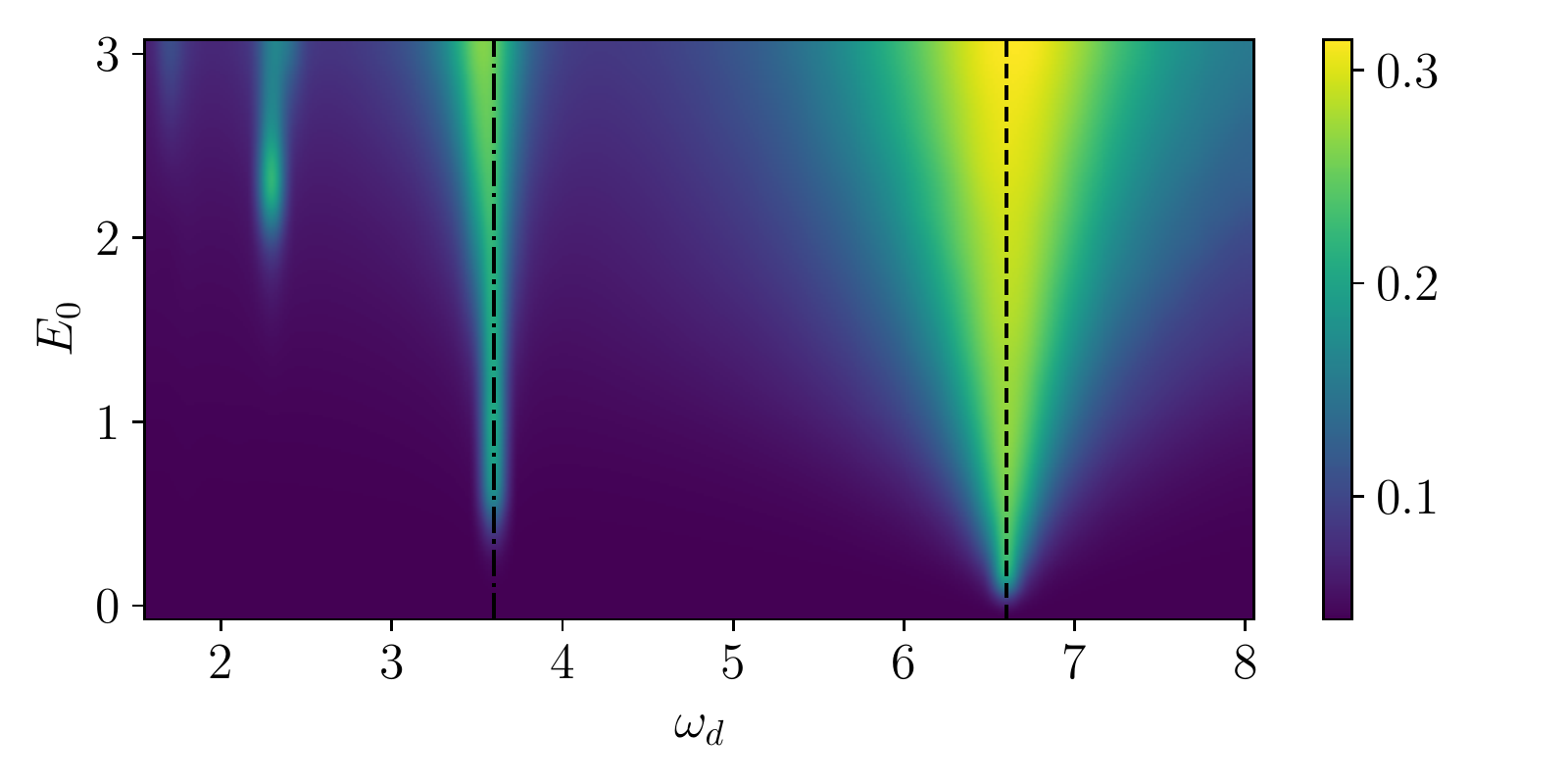}}
\caption{\label{fig:two-site_d_av} Average double occupation $\overline{d(t)}$ as a function of drive frequency $\omega_d$ and amplitude $E_0$ for the two-site Hubbard model. Time average is taken between $t = 30$ and $600$. The one- and two-photon resonances are indicated by the dashed and dash-dotted lines, respectively.}
\end{figure}

\begin{figure}
\includegraphics[width=\columnwidth]{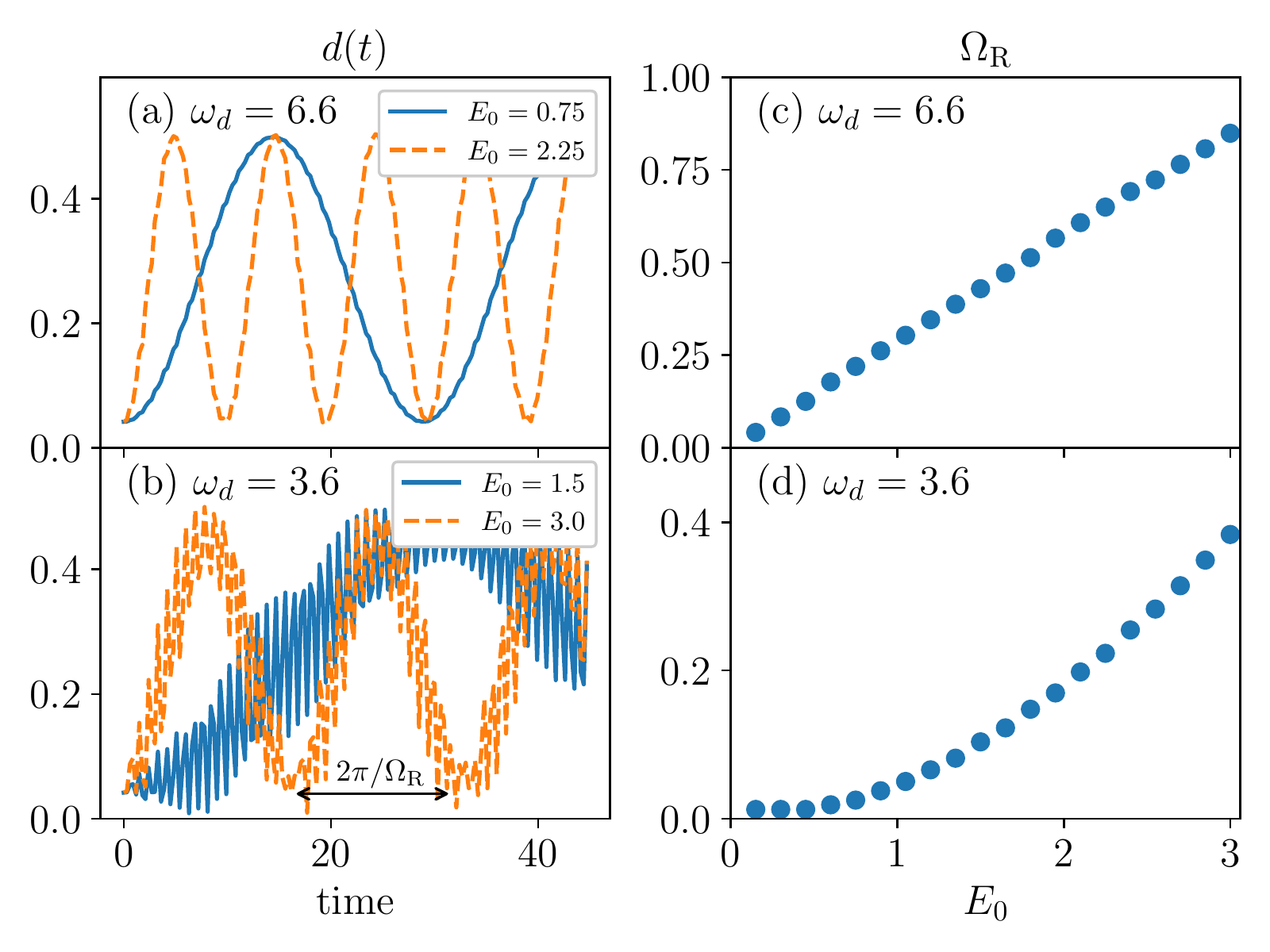}
\caption{\label{fig:two-site_Rabi} Time evolution of double occupation at the one-photon resonance $\omega_d = 6.6$ (a) and at the two-photon resonance $\omega_d=3.6$ (b) in the two-site Hubbard model. The amplitude dependence of the Rabi frequency $\Omega_R$ is linear (c) or quadratic (d), respectively.}
\end{figure}

Here we consider the two-site Hubbard model at half-filling under continuous optical excitation in order to illustrate clear Rabi oscillations at frequencies resonant with the system excitation energies and their submultiples. The Hamiltonian in Eq.~\eqref{eq:H_total} for two sites after a local gauge transformation to a purely scalar potential becomes 
\begin{multline}
H(t) = -J_0 \sum_{\sigma} \left(c^{\dagger}_{2 \sigma} c_{1 \sigma} + \text{H.c.} \right)\\
 + \sum_{i=1}^2 U n_{i\up} n_{i \dn} + \frac{E(t)}{2} (n_2 - n_1).
\label{eq:H_two}
\end{multline}
We consider the subspace with particle number $N=2$ and total spin $S=0$, which is spanned by the three states
\begin{equation}
\ket{s} = \frac{\ket{\up,\dn} - \ket{\dn,\up}}{\sqrt{2}}, \quad
\ket{D_\pm} = \frac{\ket{\up\dn,0} \pm \ket{0,\dn\up}}{\sqrt{2}}.
\end{equation}
Without the optical field, the Hamiltonian, Eq.~\eqref{eq:H_two}, is symmetric under the inversion of the two sites. Among the three states, $\ket{s}$ and $\ket{D_+}$ are parity even, while $\ket{D_-}$ is parity odd. This inversion symmetry still holds when the interaction $U$ is time dependent \cite{plodzien2019}, and different parity states are not mixed. In contrast, here the optical field breaks the inversion symmetry and we have to consider all the three states; the matrix representation of the Hamiltonian is 
\begin{equation}
H(t) = 
\begin{pmatrix}
0 & -2 J_0& 0\\
-2 J_0 & U & -E(t) \\
0 & -E(t) & U
\end{pmatrix}.
\end{equation}

At $E(t)=0$, the ground state $\ket{0}$ and the second excited state $\ket{2}$ are parity even and composed of $\ket{s}$ and $\ket{D_+}$, while the first excited state is parity odd, $\ket{1} = \ket{D_-}$. Their eigenenergies are
\begin{equation}
\epsilon_{0,2} = \frac{U}{2} \mp \frac{1}{2}\sqrt{U^2 + 16J_0^2}, \quad \epsilon_1 = U.
\end{equation}
In the eigenstate basis $\{\ket{0}, \ket{1}, \ket{2} \}$, the Hamiltonian is  
\begin{equation}
H(t) = \begin{pmatrix}
\epsilon_0 & -c_1E(t) & 0\\
-c_1E(t) & \epsilon_1 & -c_2E(t) \\
0 & -c_2E(t) & \epsilon_2
\end{pmatrix},
\label{eq:H_eigen}
\end{equation}
with $c_{1} = \braket{0|D_+}$ and $c_2 = \braket{2|D_+}$. In the Mott insulating regime, $ J_0 \ll U$, the nearly degenerate excites states are separated from the ground state by $\sim U$. The structure resembles a cascade type three-level problem and its analytical solutions with Rabi oscillations are found in Ref.~\cite{berent2010}. It can be seen from Eq.~\eqref{eq:H_eigen} that the optical field only connects the states with different parity. Thus, at first order in the field, starting from the ground state, the only possible transition is $\ket{0} \rightarrow \ket{1}$, which is resonant at $\omega_d=\epsilon_1-\epsilon_0$. At second order, on the other hand, we have transitions $\ket{0} \rightarrow \ket{1} \rightarrow \ket{2}$ resonant at $2\omega_d=\epsilon_2-\epsilon_0$. The former is a one-photon resonance and the latter is a two-photon resonance.

Starting from the unperturbed ground state $\ket{\psi(0)}=\ket{0}$ we calculate its time-evolution under the optical field $E(t) = E_0\sin\omega_d t$. In Fig.~\ref{fig:two-site_d_av}, we show the time average of the double occupation $\overline{d(t)}$ as a function of drive frequency $\omega_d$ and amplitude $E_0$. At small amplitudes, large double occupation is observed at the one-photon resonance, $\epsilon_1 - \epsilon_0 \approx 6.6$, and at the two-photon resonance, $(\epsilon_2 - \epsilon_0)/2 \approx 3.6$. 

Figs.~\ref{fig:two-site_Rabi}(a) and (b) show the time-evolution of the double occupation at these resonances. We observe nearly perfect oscillations whose frequencies $\Omega_\text{R}$ depend on the drive strength. In order to show that these stem from the Rabi oscillations, we plot the oscillation frequency $\Omega_\text{R}$ as a function of $E_0$ in Figs.~\ref{fig:two-site_Rabi}(c) and (d). At the one-photon resonance, $\omega_d = 6.6$, the oscillation frequency linearly increases with the drive amplitude $\Omega_\text{R} \propto E_0$, while at the two-photon resonance, $\omega_d = 3.6$, we find quadratic dependence $\Omega_\text{R} \propto E_0^2$. The amplitude dependence resembles that of the Rabi frequency in a two-level system \cite{fox2006}, and suggests that the observed oscillations are Rabi oscillations. At the three-photon resonance, $\omega_d \approx 2.3$, oscillating behavior appears only for strong excitation $E_0 \gtrsim 1.0$ and the relation between $\Omega_R$ and $E_0$ is obscure.

\section{Two-dimensional cluster}\label{sec:2d}
\begin{figure}
\includegraphics[width= \columnwidth]{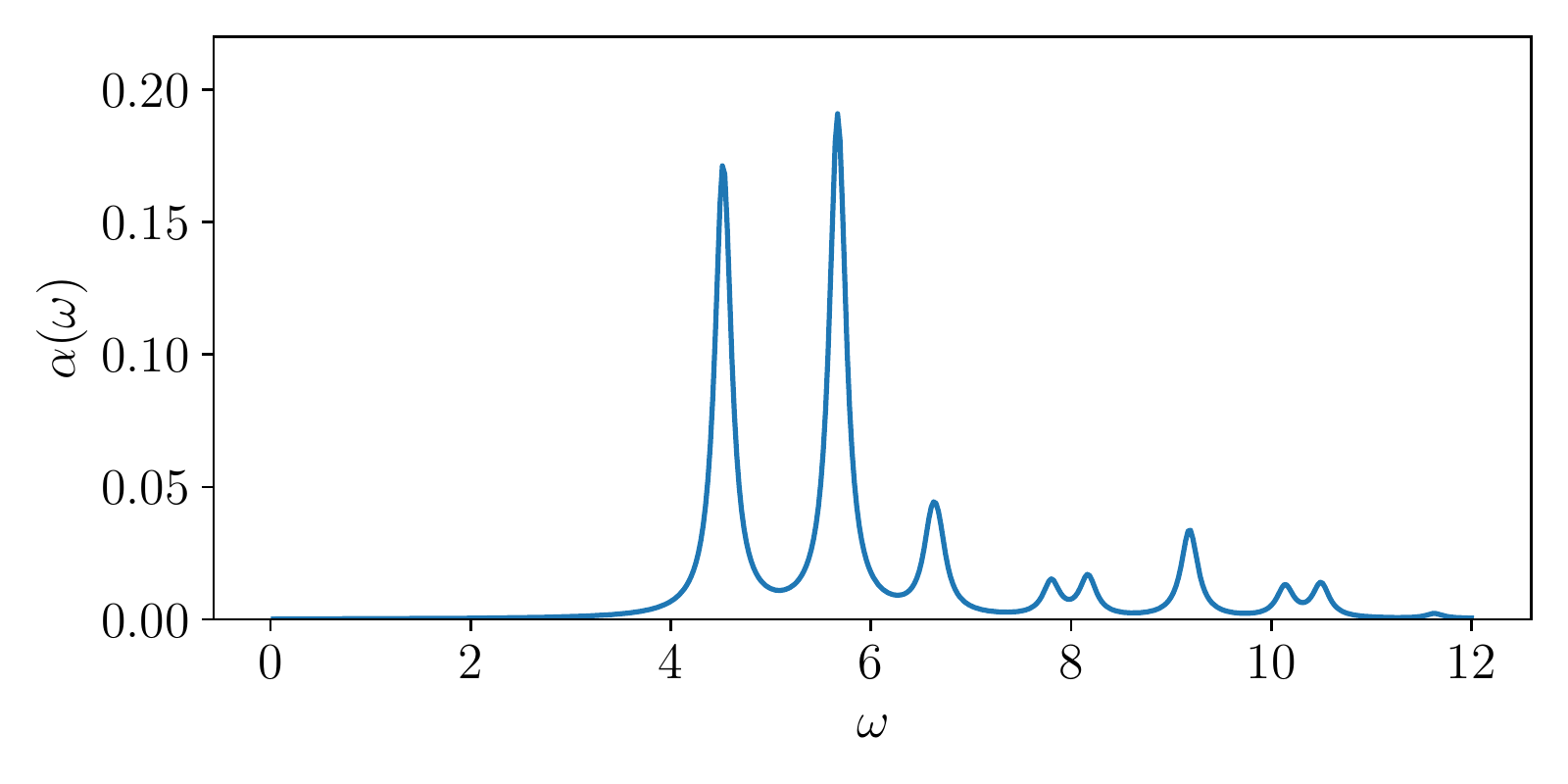}
\caption{\label{fig:alpha} Linear absorption spectrum $\alpha(\omega)$ in Eq.~\eqref{eq:alpha} for the two-dimensional case with $L=10$ sites and $U/J_0 = 6$.}
\end{figure}

Now we turn to the two-dimensional cluster with $L=10$. As a basis to interpret the time-dependent phenomena, let us first look at the linear absorption spectrum \cite{Dagotto1994, matsueda2007} 
\begin{equation}
\alpha(\omega) = - \frac{1}{\pi L^2} \Im  \bra{0} I_x \frac{1}{\omega - H + \epsilon_0} I_x \ket{0},
\label{eq:alpha}
\end{equation}
where $I_x$ is the total current operator along the $x$-axis, namely along the direction of the optical field. In Fig.~\ref{fig:alpha}, we plot $\alpha(\omega)$ for $U/J_0 = 6$ with small broadening factor $\eta = 1/L$. There are three absorption peaks around $U$ at $\omega = 4.52, 5.67$, and $6.6$, which represent transitions to the one-photon excited states.

\subsection{Double occupation}

\begin{figure}
\includegraphics[width=\columnwidth]{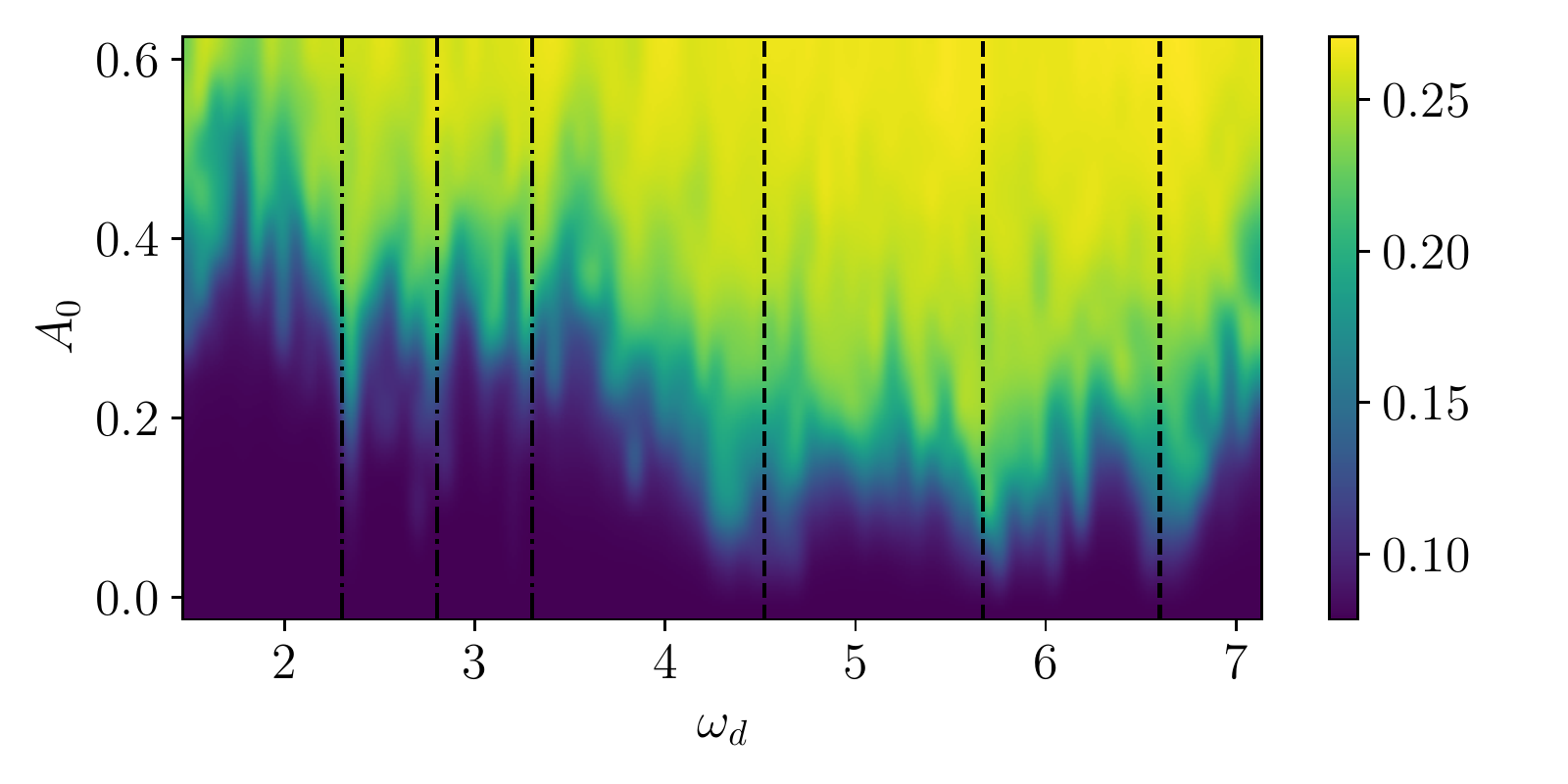}
\caption{\label{fig:2d_d_av} Average double occupation $\overline{d(t)}$ as a function of drive frequency $\omega_d$ and amplitude $A_0$ in the 2D Hubbard model. Time average is taken between $t = 30$ and $600$. The one- and two-photon resonances are indicated by the dashed and dash-dotted lines, respectively.}
\end{figure}

First, we discuss the average double occupation after the optical fields are turned on. Here the onset of the excitation is set to be $\tau_0 = 10$ with a ramp of width $\tau = 0.1$; the final steady states are almost independent of the ramp. The time-averaged values are plotted as a function of drive frequency $\omega_d$ and amplitude $A_0$ in Fig.~\ref{fig:2d_d_av}.

As expected, we find strong absorption at frequencies resonant with the three peaks in the optical absorption spectrum $\omega_d\approx4.52$, $5.67$, and $6.63$. We notice, however, that even at these resonance conditions, heating is still moderate for small field amplitudes. As in the two-site model, we find additional two-photon resonances at approximately half the frequencies of one-photon resonances. These resonances stem from second-order processes not captured by the linear absorption spectrum, Eq.~\eqref{eq:alpha}, and involve states with approximately the same parity under reflection, as we discuss below in more detail. At strong fields, the double occupation saturates to the infinite-temperature value, $d = 0.25$, at all frequencies. The appearance of the infinite-temperature state is in contrast to the two-site model, where pure oscillatory behaviors are observed due to the small dimension of the Hilbert space. More detailed discussions on the amplitude dependence are given in Appendix~\ref{sec:appA}. We also note that there is no resonance at frequencies commensurate to $U$, i.e., $\omega_d = U/n$ with $n \in \mathbb{N}$, in contrast to Ref.~\onlinecite{herrmann2017}. The discrepancy is probably due to the different lattice geometries or the effect of finite temperatures.

\begin{figure*}
\includegraphics[width=2\columnwidth]{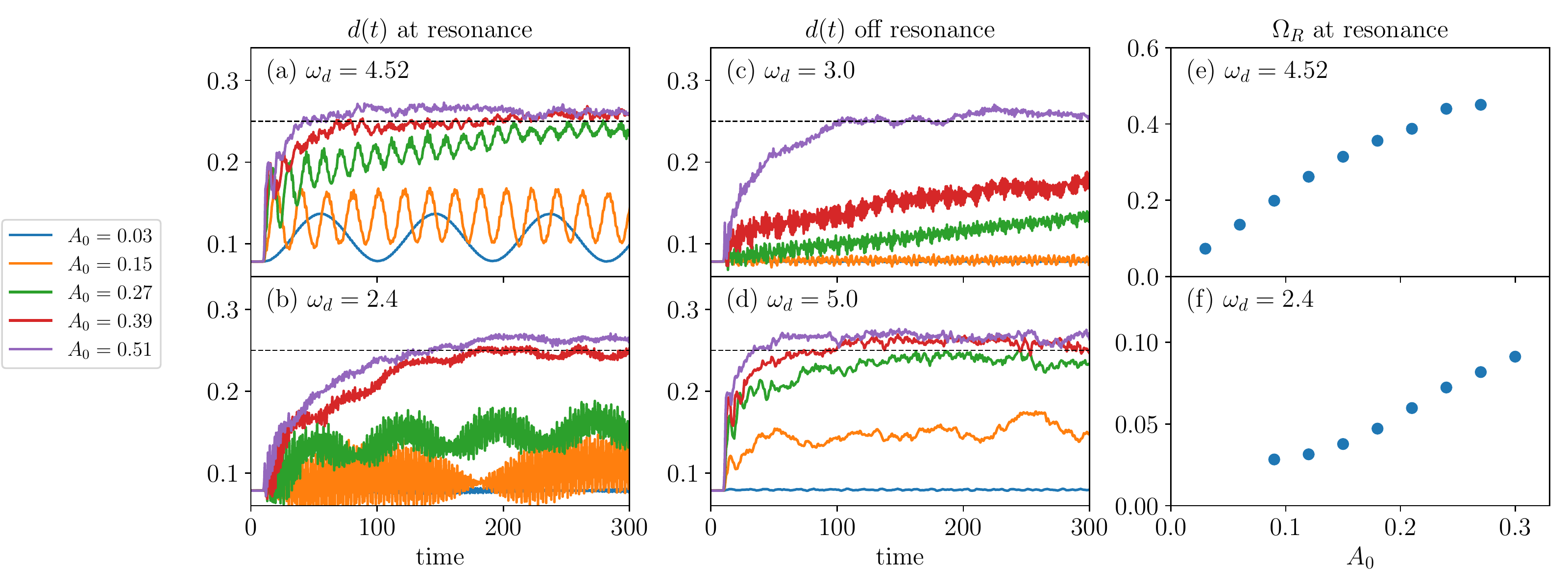}
\caption{\label{fig:2d_d_t} Time evolution of double occupation at the one-photon resonance $\omega_d = 4.52$ (a), at the two-photon resonance $\omega_d=2.4$ (b), and off resonance (c), (d) in the 2D Hubbard model. (e), (f) Amplitude dependence of the Rabi frequency $\Omega_\text{R}$ at resonance.}
\end{figure*}

In Fig.~\ref{fig:2d_d_t}(a) we plot, for various field amplitudes, the time evolution of double occupation at the one-photon resonance $\omega_d=4.52$. Similar results are obtained at the other one-photon resonances. Despite the resonance condition,  for weak fields, we find FPSs where the double occupation oscillates around a constant value. These are analogous to the Rabi oscillations in the two-site model, and they have frequencies increasing linearly with drive amplitude $A_0$ [Fig.~\ref{fig:2d_d_t}(e)]. Such a FPS with persistent oscillations is one of our main findings. In contrast, for strong fields, the oscillations are on top of an increasing value saturating around the infinite-temperature limit.

Similar oscillations are found at the two-photon resonance $\omega_d=2.40$ for weak excitation [Fig.~\ref{fig:2d_d_t}(b)], and the Rabi frequency is approximately quadratic in the field amplitude [Fig.~\ref{fig:2d_d_t}(f)], again in analogy with the two-site result. In addition to the slow Rabi oscillations, there exist fast oscillations with frequency $\omega \approx 4.5$, which presumably comes from the coherent superposition of the ground state and the excited states with absorption-peak energies. At large field amplitudes, heating and thermalization to infinite temperature occur.

Away from the resonances, Rabi oscillations disappear and two distinct behaviors arise instead. At weak fields and low frequencies [Fig.~\ref{fig:2d_d_t}(b)], the double occupation increases almost linearly in time. We may consider this linear production of doublons as an analog of the DC response \cite{oka2012, eckstein2013}. On the other hand, at weak fields and high frequencies [Fig.~\ref{fig:2d_d_t}(d)], FPSs without Rabi oscillations appear as in Refs.~\onlinecite{herrmann2017, peronaci2018}. Their steady-state values depend on the drive amplitude $A_0$ (see Appendix~\ref{sec:appA}).

\subsection{Kinetic energy}
\begin{figure}
\includegraphics[width=\columnwidth]{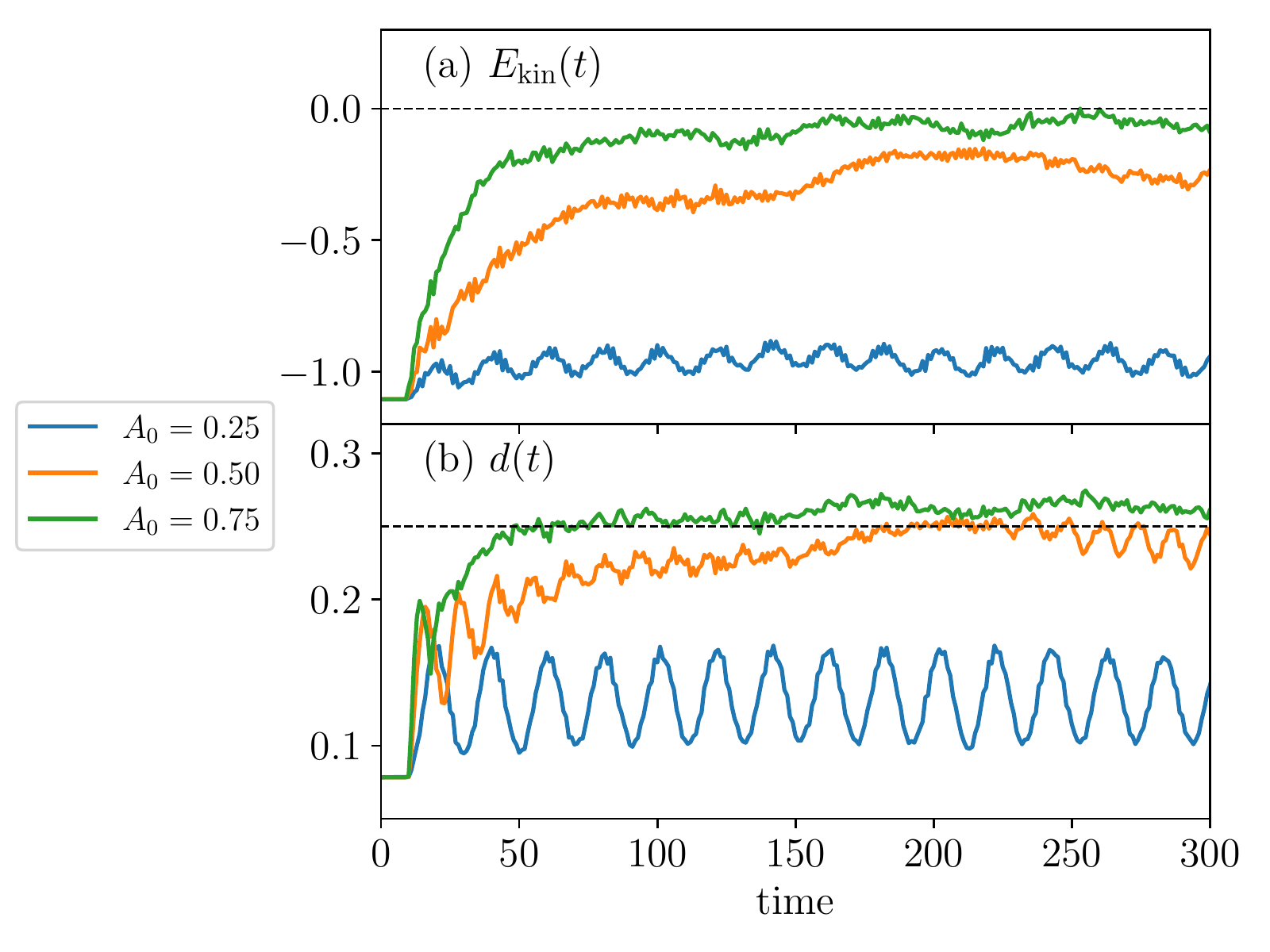}
\caption{\label{fig:2d_Ekin} Kinetic energy $E_\text{kin} (t)$ and double occupation $d(t)$ at the one-photon resonance $\omega_d = 4.52$ in the 2D Hubbard model. Dashed lines denote the infinite-temperature limit, $E_\text{kin} = 0$ and $d = 0.25$.}
\end{figure}

In addition to the double occupation, the kinetic energy also characterizes the infinite-temperature limit, where $E_\text{kin} \rightarrow 0$. In Fig.~\ref{fig:2d_Ekin}, we plot the time evolution of the kinetic energy at the one-photon resonance $\omega_d = 4.52$ for small and large field amplitudes. For comparison, we also plot the double occupation. The two observables reach steady-state values in similar time scales, which indicates that the absorbed energy is efficiently distributed between the interaction and kinetic energies. At weak excitation, Rabi oscillations appear in the kinetic energy as well, while its relative oscillation amplitude is smaller than that of the double occupation. For strong excitation, the kinetic energy goes to zero, which is consistent with the notion of the infinite-temperature limit.

\subsection{Spectral density}
In order to deepen our understanding of the Rabi oscillations and the dependence on the drive frequency and amplitude, we consider the  spectral density $S(t, \omega)$ of the time-evolved wavefunction, Eq.~\eqref{eq:spectral_func}, and the effective dimension $\kappa$, Eq.~\eqref{eq:kappa}.

\begin{figure*}
\includegraphics[width=2\columnwidth]{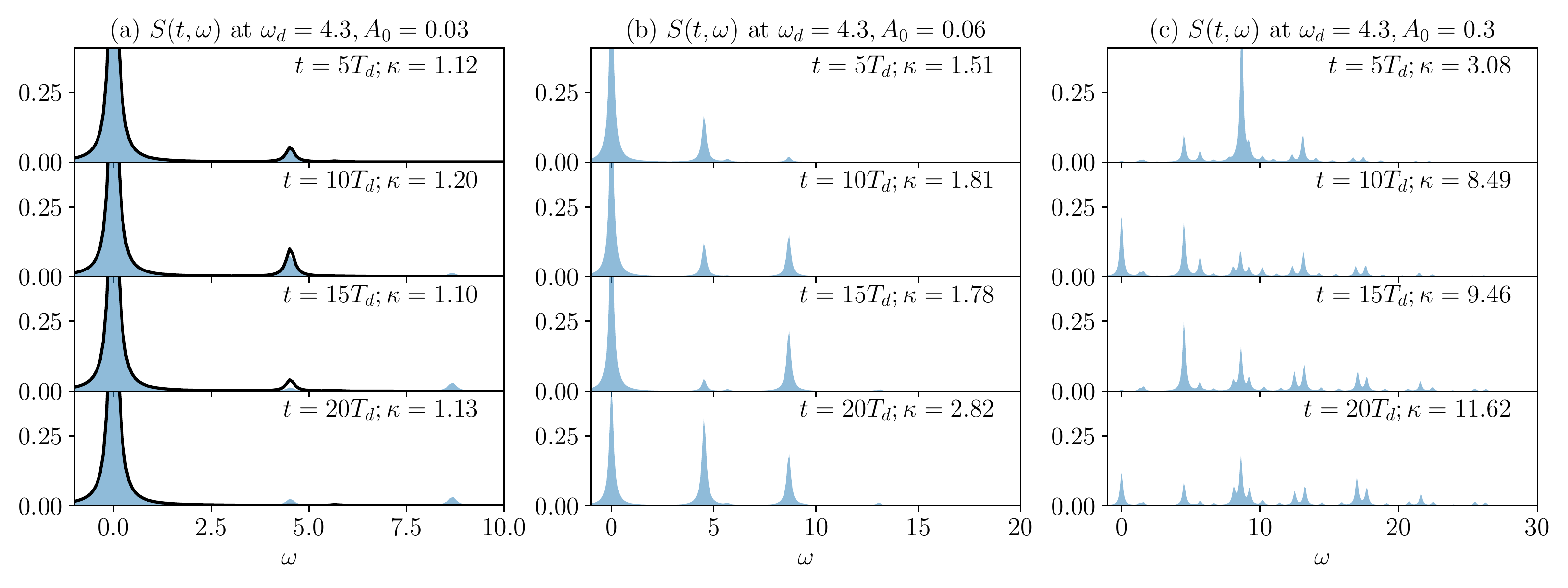}
\caption{\label{fig:spectral_density} Spectral densities $S(t, \omega)$ for frequency $\omega_d = 4.3$ and amplitude $A_0 = 0.03$ (a), $A_0 = 0.06$ (b), and $A_0 = 0.3$ (c) at times multiples of the drive period $T_d = 2\pi/\omega_d$. The delta functions are approximated by Lorentzian functions. For $A_0 = 0.03$, the approximate expression, Eq.~\eqref{eq:spectral_func_PT}, is also plotted (solid line). Effective dimensions $\kappa$ calculated from each spectrum are also shown.}
\end{figure*}

Figure~\ref{fig:spectral_density} shows the spectral density $S(t, \omega)$ at times multiple of the drive period $T_d=2\pi/\omega_d$ for drive frequency close to the one-photon resonance $\omega_d=4.3$. At small amplitude $A_0=0.03$ [Fig.~\ref{fig:spectral_density}(a)] the weight oscillates mostly between the ground state and the one-photon excited state. Thus, the effective dimension $\kappa$ remains below 2.  Increasing to $A_0=0.06$ [Fig.~\ref{fig:spectral_density}(b)], weight on the two-photon excited states $\omega \approx 2 \omega_d$ becomes not negligible, while still the weight is localized on a few states, i.e., $\kappa$ remains below 3. Finally, at strong field $A_0=0.30$ [Fig.~\ref{fig:spectral_density}(c)], the spectral density spreads over a wide energy range due to multiple-photon absorption and $\kappa > 10$, meaning that the system has reached the infinite-temperature state, i.e., the double occupation and the kinetic energy reaches the infinite-temperature values. We note that due to the limited number of Krylov dimensions (here $M=50$), the effective dimension calculated by the Lanczos method cannot reproduce the bulk infinite-temperature behavior, $\kappa \rightarrow \dim \mathbb{H}$.

\subsection{Time-dependent perturbation theory}
Here we discuss the physical origin of the Rabi oscillations at the resonant excitation and the amplitude dependence using time-dependent perturbation theory. For this purpose, we define the unperturbed zero-field Hamiltonian $H_0$ and the time-dependent perturbation,
\begin{equation}
\begin{split}
V(t) &= H(t) - H_0 \\
&= - \sum_{\braket{i,j}} J_0 \left( e^{-i \frac{e}{\hbar}  \bm{l}_{ij} \cdot \bm{A}(t)} - 1\right) c^{\dagger}_{i}c_{j} + \text{h.c.}\\
&= - I_x \sin[A(t)] - K_x \left\{ 1 - \cos[A(t)]\right\},
\end{split}
\end{equation}
where $I_x$ and $K_x$ are the current and kinetic-energy operators along the $x$-axis. Assuming weak perturbation $|A(t)| \ll 1$, we can expand $V(t)$ as \cite{Dagotto1994}
\begin{equation}
V(t) \simeq - I_x A(t)  - \frac{1}{2} [A(t)]^2 K_x.
\end{equation}
Applying first-order time-dependent perturbation theory to the ground state $\ket{0}$, we find the transition amplitude to another eigenstate of the unperturbed Hamiltonian $\ket{n}$ as
\begin{equation}
\begin{split}
\braket{n | \psi(t)} &\simeq -i \int_0^t \braket{n | V(t) | 0} e^{i \epsilon_n t'} dt' \\
&\simeq i \braket{n | I_x | 0} \int_0^t A(t') e^{i \epsilon_n t'} dt'\\
&+ \frac{i}{2} \braket{n | K_x | 0}  \int_0^t  [A(t')]^2 e^{i \epsilon_n t'} dt'
\end{split}
\label{eq:PT}
\end{equation}
The first term describes one-photon excitation and gives Rabi oscillations for $\omega_d \simeq \epsilon_n$ if the matrix element $\braket{n | I_x | 0}$ is peaked at one or a few eigenstates with similar energies. From the linear absorption spectrum, Fig.~\ref{fig:alpha}(a), we see that this matrix element has peaks, which then explains the observed Rabi oscillations. The second term, in addition to the second-order terms, describes two-photon excitation, which gives rise to the nonlinear amplitude dependence and to the increase of effective dimension by spreading occupied states.

At the lowest order in $A(t)$, the spectral density is found to be
\begin{multline}
S(t, \omega) \simeq |\braket{0|\psi(t)}|^2 \delta (\omega)\\
+ \sum_{n >0} |\braket{n | I_x| 0 }|^2 \left|\int_0^t A(t') e^{i \epsilon_{n} t'} dt' \right|^2\delta(\omega - \epsilon_{n}).
\label{eq:spectral_func_PT}
\end{multline}
Note that the ground state occupation is calculated as $|\braket{0 | \psi(t)}|^2 = 1- \sum_{n>0} |\braket{n|\psi(t)}|^2$ to avoid the use of second-order perturbation. In Fig.~\ref{fig:spectral_density}(a), the approximate expression, Eq.~\eqref{eq:spectral_func_PT}, is compared with the spectral density obtained by the Lanczos method at a weak amplitude, which well reproduces the Rabi oscillations. On the other hand, for resonant excitation, the transition probability becomes too large and makes the perturbative expression invalid. For larger amplitudes, higher orders in perturbation theory are required.

Expressions similar to Eq.~\eqref{eq:PT} can be derived for other types of time-dependent perturbation. An extensively studied example  is the driven-interaction protocol,
\begin{equation}
V(t) = \delta U(t) \sum_i n_{i\up} n_{i\dn}.
\end{equation}
In general, the matrix element $\braket{n | n_{i\up} n_{i\dn} | 0}$ is not peaked nor the excited states are degenerate, and we do not expect Rabi oscillations. However, in the limit of ${J_0/ U \ll 1}$, the excited states are nearly degenerate, and the lower and upper Hubbard bands form an effective two-level system leading to Rabi oscillations \cite{plodzien2019}.

\section{Conclusions}\label{sec:conclusions}
In this work, we have elaborated on the properties of prethermal Floquet states in optically excited Hubbard clusters. We have demonstrated that the prethermal states exist even at resonance, as far as the drive strength is weak. In particular, the Floquet prethermal states at resonance involve Rabi oscillations, whose frequency scales linearly with the field amplitude at one-photon resonances, and with the square amplitude at two-photon resonances. We have elucidated the origin of the Rabi oscillations by time-dependent perturbation theory and with an analysis of the optically-driven two-site Hubbard model.

The observation of Rabi oscillations in this model suggests possible coherent manipulation of quantum many-body states. Rabi-like oscillations are expected to be a general phenomenon for resonantly driven small Hubbard-like clusters; for example, similar oscillations are found after short pulses \cite{okamoto2019, tanaka2020}. Considering the coherent nature of Rabi oscillations, the mechanism can be combined with different shapes of drive pulses and used for optical control of correlated quantum states. While we expect that introducing dissipation to our model will weaken the Rabi oscillations, it is an interesting open question if the oscillations survive or completely vanish.

\acknowledgments
This work is supported by Research Foundation for Opto-Science and Technology and by Georg H. Endress Foundation. The authors acknowledge support by the state of Baden-W\"urttemberg through bwHPC. We thank S. Stumper for constructive discussions on the manuscript.

\appendix
\section{Amplitude dependence of the steady states}\label{sec:appA}

\begin{figure}
\includegraphics[width=\columnwidth]{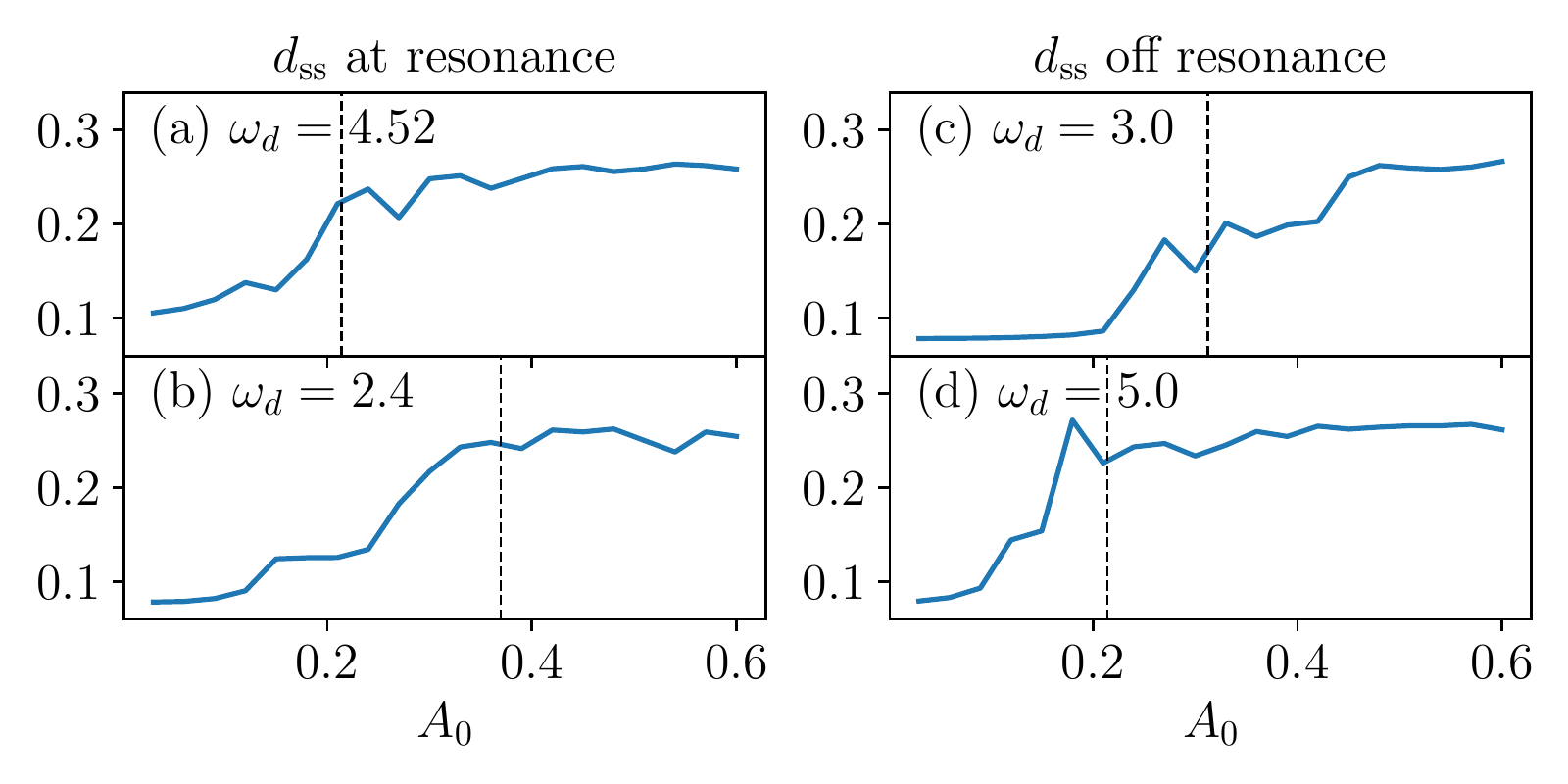}
\caption{\label{fig:2d_d_A0} Steady-state values of the double occupation $d_\text{ss}$ at the one-photon resonance $\omega_d = 4.52$ (a), at the two-photon resonance $\omega_d = 2/4$, and off resonance (c), (d) in the 2D Hubbard model. Time average is taken between $t = 550$ and $600$. Dashed lines indicate the drive amplitude where the effective dimension becomes $\kappa \approx 1.5$ after the one cycle of the pulse, $t = T_d$, which approximately separate the FPS and the infinite-temperature state.}
\end{figure}
\begin{figure}[!t]
\includegraphics[width=\columnwidth]{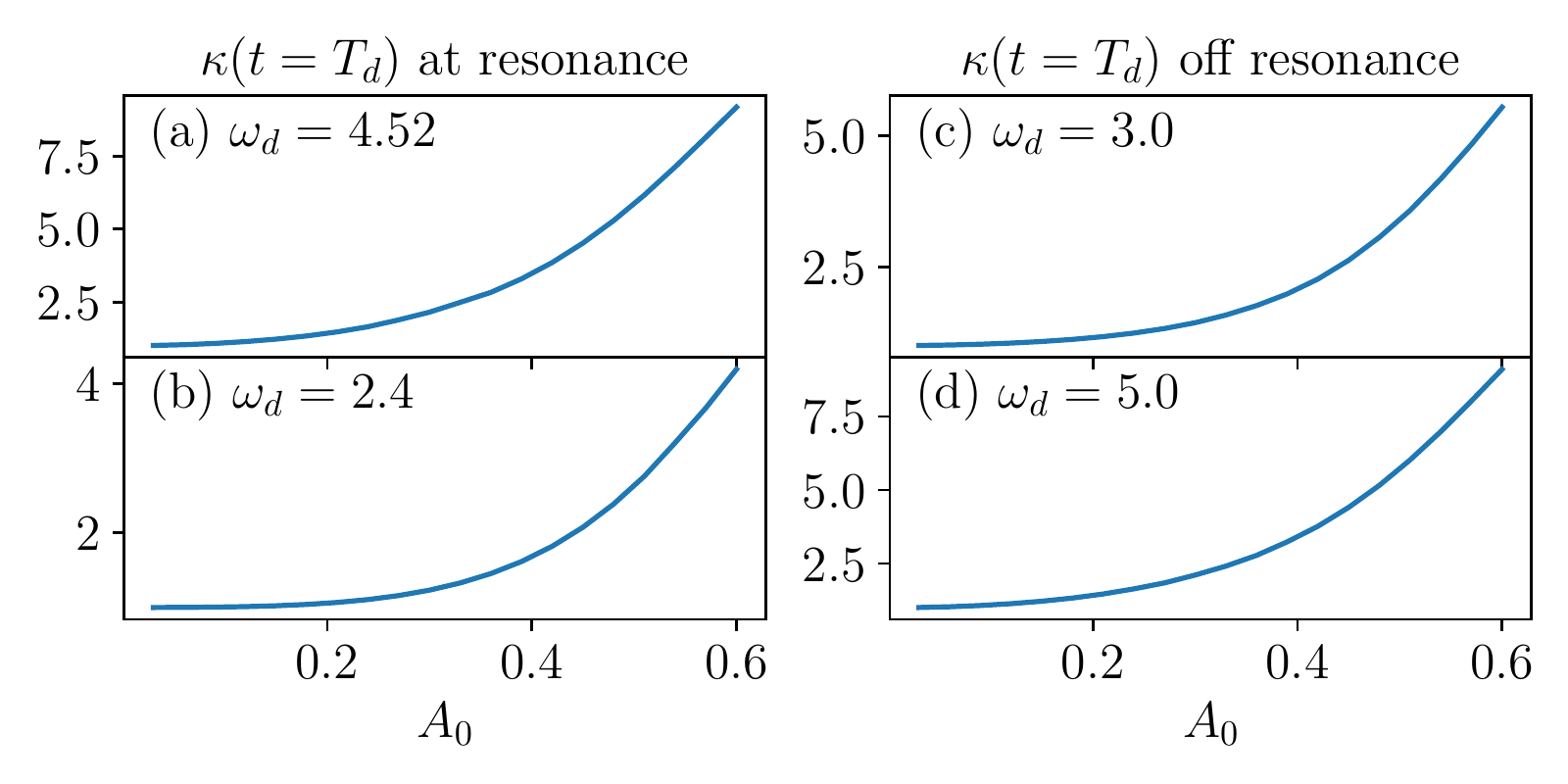}
\caption{\label{fig:2d_deff_A0} Effective dimensions $\kappa$ after the one cycle of the pulse, $t = T_d$ at the one-photon resonance $\omega_d = 4.52$ (a), at the two-photon resonance $\omega_d = 2.4$, and off resonance (c), (d) in the 2D Hubbard model. }
\end{figure}

In this appendix, we give details on the amplitude dependence of the steady states. Figure~\ref{fig:2d_d_A0} shows the amplitude dependence of the steady-state values of the double occupation at resonance and off resonance. We see that steady states depend on the drive strength nonlinearly. At resonance, the double occupation grows almost quadratically as $A_0$ increases and reaches the infinite-temperature value, $d=0.25$. On the other hand, off resonance, the change of the double occupation is more stepwise. It is an open question if the transition becomes sharper as we increase the system size, indicating a nonequilibrium phase transition. The perturbative expression in Eq.~\eqref{eq:PT} explains the nonlinear dependence for weak excitation, while higher-order calculations are necessary to understand the transition to the infinite-temperature state.

We now focus on the states at $t = T_d = 2\pi /\omega_d$, i.e., after one cycle of the optical excitation. The effective dimensions $\kappa(T_d)$ are plotted in Fig.~\ref{fig:2d_deff_A0}. As expected, the effective dimension increases with the drive amplitude $A_0$. In particular, compared with Fig.~\ref{fig:2d_d_A0}, for cases with $\kappa \gtrsim 1.5$, the system evolves further to reach the infinite-temperature state. For $\kappa < 2$, the wavefunction is dominated by the ground state and the one-photon excited state. In this case, further application of the optical pulses does not spread the wavefunction to higher excited states, but rather shows Rabi oscillations.

\end{document}